\documentstyle[aps,prl,multicol,epsf]{revtex}

\begin{document}

\title{Fast coarsening in unstable epitaxy with desorption}
\author{P.\ \v{S}milauer,$^1$ M.\ Rost,$^2$ and J.\ Krug$^3$}
\address{
$^{(1)}$Institute of Physics, Academy of Sciences of the Czech Republic,
Cukrovarnick\'{a}~10, 162~53~Praha~6, Czech Republic\\
$^{(2)}$Helsinki Institute of Physics, P.L.\ 9, 00014 University of Helsinki,
Finland\\
$^{(3)}$Fachbereich Physik, Universit\"at GH Essen, 45117 Essen, Germany
}
\date{\today}
\maketitle
\begin{abstract}
Homoepitaxial growth is unstable
towards the formation of pyramidal mounds when interlayer
transport is reduced due to activation barriers to hopping at step edges.
Simulations of a lattice model and a continuum equation
show that a small amount of desorption dramatically speeds
up the coarsening of the mound array, leading
to coarsening exponents between 1/3 and 1/2.
The underlying mechanism is the faster growth of larger
mounds due to their lower evaporation rate.
\end{abstract}

\begin{multicols}{2}

Growth of perfectly smooth epitaxial layers, e.g.,
for the fabrication of two-dimensional electron-gas heterostructures,
requires suppressing growth instabilities that lead to
surface roughening. However, one can also try to harness the
instabilities and make them produce regular arrays of tiny objects on
the crystalline substrate, such as quantum wires and quantum dots.
Much attention has been paid to nanostructures created during
heteroepitaxial growth. It seems tempting to use also other
types of instabilities associated with {\it homoepitaxy\/}.

The most prominent of these kinetic instabilities is the creation of pyramidal
features on the surface as a result of additional activation
barriers to hopping at step edges reducing
interlayer transport \cite{villain1,johnson}.
Theoretical \cite{siegert,SV,AF,golub,krug,stroscio} and experimental
\cite{stroscio,moundexp} research has shown that
the array of pyramidal mounds coarsens in time with the average mound
size $\xi(t)$ increasing according to a power law,
$\xi \propto t^{1/z}$.
Under the standard assumption of negligible desorption, which
implies volume conservation for the growing film \cite{villain1},
the exponent $1/z$ was demonstrated to have an upper bound of 1/4
\cite{rost1,note:anis}. Here we show, using kinetic Monte Carlo (KMC)
simulations
and numerical integration of a continuum equation of motion,
that even a minute amount of desorption
drastically changes this situation with $1/z$ increasing towards
1/2. Detailed investigation of growth kinetics reveals that
the reason is a dependence of the evaporation rate on the mound size, leading
to faster growth of big mounds at the expense of small ones.

\noindent
{\em Monte-Carlo simulations.}
To study epitaxial growth with desorption, we used a solid-on-solid
KMC model in which the crystal is
assumed to have a simple cubic structure with neither bulk vacancies
nor overhangs allowed.  The basic processes included in our
model are the deposition of atoms onto
the surface at a rate $F$, their surface diffusion, and evaporation
from the surface.  The diffusion of surface adatoms is
modeled as a nearest-neighbor hopping process at the
rate $k_D$ $=$ $k_0\exp(-E_D/k_{B}T)$, where $E_D$ is the hopping barrier,
 $T$ is the substrate temperature, and $k_B$ is Boltzmann's constant. The
pre-factor $k_0$  is the attempt frequency of a surface adatom and is
assigned the value $10^{13}$~s$^{-1}$.
The barrier to hopping is given by
$E_D = E_S + nE_N + (n_i-n_f)\Theta(n_i-n_f)E_B$
where $E_S$, $E_N$, and $E_B$ are model parameters, $n$
is the number of in-plane nearest neighbors before
the hop, $n_i$ and $n_f$ are the number of next-nearest neighbors in the
planes beneath and above the hopping atom before ($n_i$) and after ($n_f$)
a hop, and $\Theta(x)$=1 if $x>0$, and 0 otherwise
(cf. Ref.~\onlinecite{SV} for a more detailed description of the model).
The evaporation of a surface adatom occurs at the rate
$k_{\rm ev}=k_0\exp(-E_{\rm ev}/k_BT)$, where $E_{\rm ev}=E_0 + nE_N$
with $E_0$ being the energy for evaporation of a {\rm free\/} surface adatom.

\begin{figure}[htbp]
\narrowtext
\unitlength1cm
\begin{center}
   \begin{picture}(8,6.0)
      \includegraphics{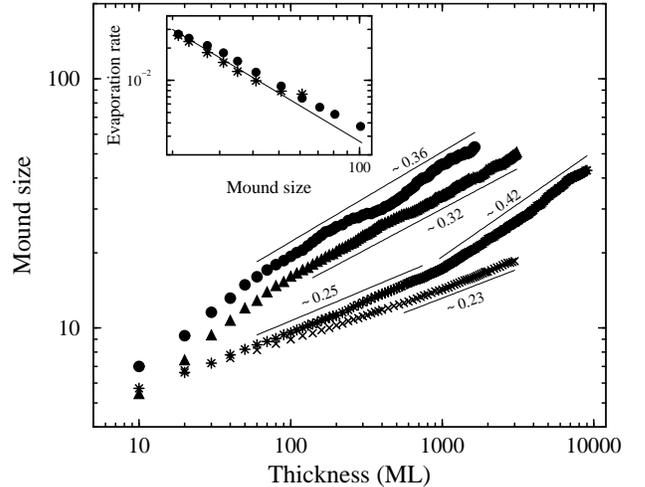}
   \end{picture}
\end{center}
\caption{Lateral mound size $\xi(t)$ evolution for different model
parameters and growth conditions obtained in KMC simulations.
The model parameters used were
$E_0$$=$$1.9$ eV at $T$$=$$750$ K, $R_i$$=$$0$ (stars) and $R_i$$=$$3$
(circles), $E_0$$=$$2.0$ eV at $T$$=$$750$ K, $R_i$$=$$3$ (crosses),
and $E_0$$=$$1.75$ eV at $T$$=$$670$ K, $R_i$$=$$3$ (triangles).
The slopes indicated are results of least-squares fitting. The
error bars of the exponents (estimated from run-to-run variations)
are of the order of 0.01.
Inset:
Size-dependent contribution to the evaporation rate of a single mound
determined on lattices of size 21, 23, 27, 31, 35, 41, 51, 61,
71, 81, and 101 with periodic boundary conditions. Data show
average over 25 runs, in each of which about
1000 ML were deposited. The full line has the form $\Delta v =
A/\xi^{1.5}$, and is consistent with coarsening exponents
for the same model parameters.}
\label{fig:KMC}
\end{figure}

The simulations were carried out on square 300$\times$300 to
600$\times$600 lattices  with periodic boundary conditions.
The basic set of model parameters and growth
conditions used was:
$E_{S}=1.54$~eV, $E_{N}=0.23$~eV, $E_B=0.175$~eV,
and $F$=1/6 monolayer (ML)/s (set I of Ref.~\onlinecite{SV}).
Under these conditions the equilibrium evaporation flux
$F_{\rm eq} = k_0 \exp[-(E_0 + 2 E_N)/k_B T)]
\approx (10^{-3} - 10^{-2}) \times F$,
and the actual desorption rate is a few times larger.
The robustness of the observed
behavior was tested by using different temperatures and
evaporation barriers $E_0$,
and by including the ``incorporation radius'' effect whereby
the incoming atom is placed at the site with the highest number of
lateral nearest neighbors within a square area of size $R_i$
centered on the site of incidence \cite{SV}.

Simulation results are shown in Fig.~\ref{fig:KMC} \cite{note:correlation}.
As the desorption
rate increases, coarsening becomes faster and the coarsening exponent
$1/z$ becomes much bigger than 0.19-0.26, the range of values observed
in previous simulation work using the same model without desorption
\cite{SV}. Even a rather small amount of desorption
\cite{strong} thus drastically affects
the coarsening exponent regardless of the details of the simulation
model (such as the model parameters and the incorporation radius).
In the regions of fits shown in Fig.~\ref{fig:KMC} the mound slope
stays approximately constant, indicating that the asymptotic
regime has been reached.
An interesting feature of Fig.~\ref{fig:KMC} is the crossover
observed for the case $R_i$ $=$ $0$ after approx. 1000~ML were deposited.
We discuss the underlying change in the mechanism of coarsening below.

\begin{figure}[htbp]
\unitlength1cm
\begin{center}
   \begin{picture}(8,4)
      \includegraphics{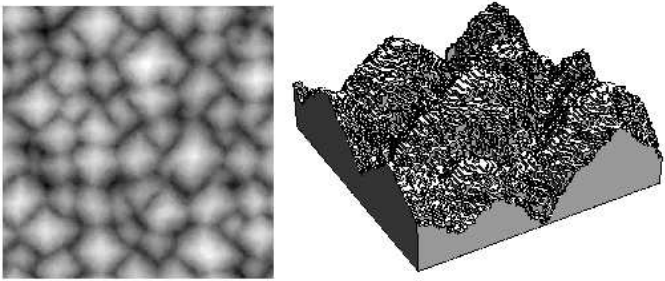}
   \end{picture}
\end{center}
\caption{Surface morphology in KMC simulations
after approx.\ 2000~ML have been deposited.
The displayed part of the lattice is 300$\times$300
(plane view, left) and 100$\times$100 (perspective plot, right).}
\label{fig:maps}
\end{figure}

Fig.~\ref{fig:maps} shows plane and perspective views of the surface
morphology after approx. 2000~ML have been deposited. Pyramidal
mounds are separated by narrow, deep troughs. The surface profile
is clearly asymmetric with flat, rounded mound tops and sharp,
deep valleys. For the case of $R_i$ $=$ $3$, mounds are much shallower
and bigger, having also more regular structure. In both cases,
however, very fast coarsening is observed.

\noindent
{\em Continuum equations.}
Further confirmation of the dramatic effect of evaporation comes
from continuum theory, where the surface is modeled by a smooth, space
and time-dependent height function $H({\bf x},t)$.
Our starting point is the standard continuum equation for unstable
epitaxy \cite{krug,rost1}, to which the leading order effect of desorption
is added in terms of a slope-dependent growth rate $V(\vert \nabla H
\vert)$:
\begin{equation}
\label{cont1}
\frac{\partial H}{\partial t} = - K \Delta^2 H- \nabla \! \cdot \!
[f(\vert \nabla H \vert^2) \nabla H] + V(\vert \nabla H \vert).
\end{equation}
In the absence of desorption $V(u) \equiv F$, the external flux.
For a vicinal surface  with step spacing $\ell$ (tilt $u = a/\ell$, with
lattice constant $a$), the growth rate according to
BCF theory reads \cite{bcf}
$
V_{\rm BCF}(\ell) = (F x_s/\ell) \tanh(\ell/x_s)
$
where $x_s = 2 \sqrt{D \tau}$ is the desorption length, depending on
the diffusion coefficient $D$ and the desorption rate $1/\tau$ from a flat
surface.

To use the BCF-expression also for near singular surfaces \cite{pimp},
we introduce an effective, tilt-dependent step spacing $\ell_{\rm
eff}$, which equals $\ell$ in the step flow regime, $u \gg a/\ell_D$,
and reduces to the terrace size or island distance \cite{villain2}
$\ell_D$ for small $u$. Desorption is considered a small, perturbative
effect in the sense that
\begin{equation}
\label{condition}
\alpha \equiv \ell_D/x_s \ll 1,
\end{equation}
which means that it is much more likely for an atom to be captured at
a step than to desorb. Under this
condition the terrace size $\ell_D$ should not be influenced by desorption.
A plausible formula for $\ell_{\rm eff}$ is $\ell_{\rm eff}(u)$ $=$
$\ell_D [1$ $+$ $u^2 (\ell_D/a)^2]^{-1/2}$, and the growth rate is then
$V(u) = V_{\rm BCF}(\ell_{\rm eff}(u))$.
Because of (\ref{condition}), $\ell_{\rm eff}(u) \leq \ell_D \ll x_s$
for all slopes, and therefore we can expand $V_{\rm BCF}$ to obtain
\begin{equation}
\label{Vapprox1}
V(u) \approx
F[ 1 -  (\alpha^2/3)(1 + u^2 (\ell_D/a)^2)^{-1}].
\end{equation}
The growth rate varies between $V(0) = F[1 - (1/3) \alpha^2]$ for the
singular surface and $F$ for $u \gg a/\ell_D$.

In the second term on the right hand side of
Eq.\ (\ref{cont1}) we set $f(u^2)$ $=$
$f_0 [1 \!- \!(u/m_0)^2]$, which leads to a
stable selected slope $m_0$ \cite{siegert}
as is observed in our
lattice simulations \cite{nostableslope}.
We subtract the deposited film thickness, $H \to H - Ft$ and rescale
time, lateral space, and height variables to
arrive at the dimensionless form \cite{rost1}
\begin{equation}
\label{cont2}
\frac{\partial h}{\partial t} = - \Delta^2 h - \nabla \! \cdot \!
\Bigl[\Bigl(1 \! - \! (\nabla h)^2 \! \Bigr) \; \! \nabla h \Bigr] -
\frac{\alpha^2/3}{1 + (\nabla h)^2}.
\end{equation}
The one-dimensional version of (\ref{cont2}) with an evaporation
rate $\sim (\nabla h)^2$ was considered in a different
context by Emmott and Bray \cite{bray}.

\begin{figure}[htbp]
\unitlength1cm
\begin{center}
   \begin{picture}(8,6)
      \includegraphics{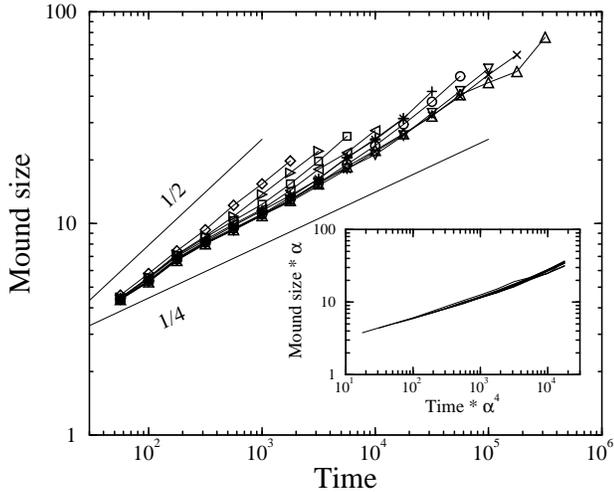}
   \end{picture}
\end{center}
\caption{Lateral mound size $\xi(t)$ for nine different evaporation strengths
$\alpha^2/3 \! = \! 10^{-1/2}$, $10^{-5/8}$, $10^{-3/4}$, \dots, $10^{-3/2}$
obtained by numerical integration of the continuum equation.
The transient regime $\xi \! \sim \! t^{1/4}$ persists until
evaporation dominates the surface driven coarsening at $t_\alpha \!
\sim \! \alpha^{-4}$, the crossover time to asymptotic fast coarsening
$\xi \! \sim \! t^{1/2}$.
Scaling plot in the inset shows time $t \times \alpha^4$ and length
$\xi \times \alpha$.}
\label{xi_alpha}
\end{figure}

We integrated Eq.\ (\ref{cont2}) numerically (for the method and system
sizes see \cite{RSK1}) and found similar behavior as in the lattice model.
After an initial fast increase of the lateral mound size $\xi$,
the pattern coarsens as in the case without
evaporation, $\xi \sim w \sim t^{1/4}$ \cite{rost1}
($w$ is the mean square width of the surface,
 $w^2 = \langle {\tilde h}^2 \rangle$ where $\tilde h \! = \! h \! - \!
\langle h \rangle$ denotes the height profile relative to its mean).
This behavior is transient and eventually crosses over to a fast
asymptotic increase of
the mound size and the surface width as $\xi \sim w \sim t^{1/2}$. The
mound size $\xi(t) $ is shown in Figure \ref{xi_alpha} for values of
$\alpha^2/3 $ ranging from $10^{-1/2}$ to $10^{-3/2}$, decreased by a
factor $10^{-1/8}$ between succeeding curves.
The transient $t^{1/4}$--regime is absent
for the strongest evaporation $(\alpha^2/3 \! = \! 10^{-1/2})$ and
becomes more pronounced as $\alpha$ is decreased. A similar
crossover is observed in the KMC simulations with $R_i \! = \! 0$
(stars in Fig. \ref{fig:KMC}).

\begin{figure}[htbp]
\unitlength1cm
\begin{center}
   \begin{picture}(8,4)
      \includegraphics{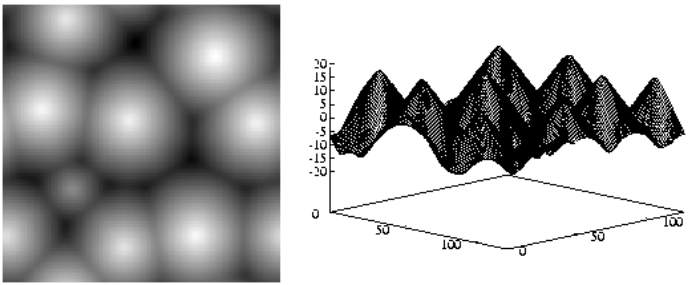}
   \end{picture}
\end{center}
\caption{Surface profile from continuum equation
(evaporation $\alpha^2/3$$=$$10^{-1/2}$)
at late times when $\xi \! \sim \! t^{1/2}$. Conical mounds
(right) form a cellular structure visible in the greyscale representation
(left).}
\label{cones}
\end{figure}

The evaporation term in Eq.\ (\ref{cont2}) breaks the up--down symmetry
($h \leftrightarrow \! -h$)
\cite{note:sym}. When it is dominant (in the fast coarsening
regime at late times) the surface morphology is asymmetric. The profile
shown in Figure \ref{cones} consists of conical mounds
separated by well defined, narrow valleys. Notice that there are no
``negative mounds''. The greyscale plot shows
the cellular arrangement of the cones. The observed
features are
very similar to results of simulations in Fig.~\ref{fig:maps}.

\noindent {\em The origin of fast coarsening.\/}
Allowing for evaporation fundamentally changes the nature of
the growth instability, because it introduces a coupling between
the local growth rate and the surface morphology. In particular,
one expects generically a dependence of the average growth rate of a mound
on its size, while for conserved growth only the {\em fluctuations}
in the growth rate are size dependent \cite{SmilVvedTang}. To see
how this can affect the coarsening law,
assume that the growth
velocity of a mound depends on its size $\xi$ as
\begin{equation}
\label{size}
v(\xi) = v_0 - \Delta v(\xi) \; \; \; \mbox{where} \; \; \;
\Delta v(\xi) \sim \xi^{-\nu}
\end{equation}
with a {\em positive} prefactor, so that large mounds grow faster.
If $\xi$ is the only macroscopic length in the system, the size differences
between mounds are also of the order of $\xi$. The time scale on which
a small mound is eliminated by its larger neighbors is then given by
$t_{\xi} \! \sim \! w/\Delta v(\xi)$, since the surface
width $w$ equals the typical height of mounds.
Using that the mounds have a constant slope,
$w \sim \xi$, it follows that $\xi \sim t^{1/(1+\nu)}$
or $z = 1 + \nu$. Provided $\nu < 3$ this violates
the bound $z \geq 4$ obtained in the
conserved case \cite{rost1}.

A well-known mechanism for a size-dependent growth rate
is the Gibbs-Thomson-effect: For spherical
droplets in equilibrium
the evaporation rate is proportional to the curvature $\sim
1/\xi$, hence within the Wilson-Frenkel-approximation \cite{gilmer}
the growth rate is of the form (\ref{size}) with
$\nu = 1$. The mounds in our
lattice simulations
are more conical in shape, with rather straight sides and
rounded regions of lateral extent $\approx \ell_D$ at the tips. Assuming
that desorption occurs preferentially
from the tip regions, the evaporation rate
of a mound of size $\xi$ has a contribution
proportional to the ratio of the tip area $\sim \ell_D^2$
to the mound area $\sim \xi^2$,
leading to $\nu = 2$ and $1/z = 1/3$. For a more quantitative estimate,
$\Delta v(\xi)$ was determined in a sequence of
simulations on small square lattices of lateral
size 21, 23, \dots up to 101 \cite{note:range}.
As initial configuration on each
of them a single mound was prepared. It persisted during deposition of
1000 monolayers, and the average evaporation rate was determined from
a sequence of 25 runs. The data presented in Figure
\ref{fig:KMC} show that $\Delta v(\xi) > 0$ and
$\nu = 1.5 \pm 0.1$, which is consistent with direct observations
of coarsening on large lattices (cf. Fig.\ \ref{fig:KMC}).

An analytical evaluation of $\Delta v(\xi)$ is possible for the
continuum equation, which will also allow us to derive the scaling of
the crossover times (cf. Fig. \ref{xi_alpha}) with $\alpha$. We recall
the surface profile of Figure \ref{cones}. The cones show {\em two\/}
lateral lengthscales: (i) their size $\xi$, which for late times is
much larger than (ii) $\ell_D$, the diameter of the tips and the
valleys ($= \! O(1)$ in our rescaled units) which is independent of
$\xi$.

Thus for a mound on a $d$--dimensional surface the fraction of the
surface covered by the tip is $(1/\xi)^d$. The surrounding trough has
codimension one and a relative weight $1/\xi$, while the major part of
the surface consists of the sloped sides of the conical mounds.

Evaporation is less pronounced on the mounds' sides whereas it is
enhanced by an amount of order $\alpha^2 $ on the horizontal parts,
i.e.\ on the tip and in the surrounding valley. As a consequence
equation (\ref{size}) also holds for the continuum equation, where
$v_0$ is the evaporation rate on the mounds' sides, and the enhanced
mass loss from small mounds is mainly due to the surrounding valley,
i.e.\  $\Delta v(\xi) \sim \alpha^2 / \xi $. So the timescale for mound
coalescence is $t_{\xi} \! \sim \! w/\Delta v(\xi) \! \sim \!
\xi^2/\alpha^2$ (due to the stable slope, $w \! \sim \! \xi$), and it follows
that $\xi \sim \alpha t^{1/2}$. Incidentally, the
same coarsening law was found in the one-dimensional case \cite{bray}.

The initial increase $\xi \sim t^{1/4}$ is not due to evaporation and
thus the same for all values of $\alpha$ (see Figure
\ref{xi_alpha}). Together with the late time behavior
$\xi \sim \alpha t^{1/2}$
this yields the estimate $t_\alpha \! \sim \! \alpha^{-4} = (x_s/\ell_D)^4$
for the time at which evaporation begins to dominate the coarsening
process.
Rescaling time as $t/t_\alpha$ and length as $\xi / t_\alpha^{1/4}$,
and omitting the initial fast increase puts all curves and in
particular the crossover times $t_\alpha$ on top of each other,
as shown in the inset of Figure \ref{xi_alpha}.

We emphasize that fast coarsening
for the continuum equation is due to the dominance of evaporation
from valleys compared to the rest of the mounds. Direct inspection
shows that in the KMC simulations more atoms in fact evaporate from
the {\em upper\/} parts of the mounds. This explains why the
coarsening exponent observed for the lattice model is smaller than
1/2: Given enhanced evaporation only on the tips, $\Delta v (\xi)$ is
of the order of $\alpha^2/\xi^d$, leading to
$\xi \sim (\alpha^2 t)^{1/(d+1)}$ in $d$ dimensions, hence $z = 3$ for
$d=2$ as argued previously. To improve on this estimate, more detailed
information about the shape of mounds and its coupling to the evaporation
rate would be needed. It is nevertheless interesting to note that the
coarsening exponent $1/z = 1/(d+1)$ is always larger than the value
$1/z = 1/(d+2)$ obtained for noise-induced coarsening \cite{SmilVvedTang},
indicating that our conclusions will not be modified by shot noise.

In summary, we have identified a general mechanism for fast
mound coarsening in unstable growth with desorption. While the detailed
appearance of the effect is different in the lattice model as compared
to the continuum equation, in both cases the key feature is the dependence
of growth rate on mound size. This gives us confidence that the phenomenon
is robust and will be observed under suitable experimental conditions.

\noindent
{\em Acknowledgements.} Support by Volkswagen-Stiftung (P.\v{S}., J.K.),
DFG/SFB 237 (J.K., M.R.), the Grant Agency of the Czech Republic,
Grant No. 202/96/1736 (P.\v{S}.)
and the COSA program of the Academy of Finland (M.R.)
is gratefully acknowledged.

\end{multicols}


\begin{references}
\bibitem{note:correlation} The mound size $\xi(t)$ is measured as the
full width at half maximum of the height-height correlation function
\cite{RSK1}.
\bibitem{nostableslope} Fast coarsening with desorption does not 
require the presence of a stable slope $m_0$. We also observed it 
for other forms of $f$  introduced in \cite{rost1}, in particular
$f(u^2) \! = \! f_0/(1 \! + \! u^{\gamma+1})$ with 
$\gamma \! = \! 1$ and $3$.
\bibitem{note:sym} Conservative terms breaking the up/down symmetry
are also expected to be present \cite{krug,stroscio}, however since they
do not affect the coarsening law \cite{stroscio,paolo} they will be
ignored here.
\bibitem{note:range} This is more than the range which is covered by the mound
size during our growth simulations.
\end{references}
\end{document}